\documentclass[aps, twocolumn,superscriptaddress,floatfix]{revtex4-2}

\usepackage{dcolumn}
\usepackage{bm}

\usepackage{relsize}
\usepackage[utf8]{inputenc}
\usepackage[english]{babel}
\usepackage[T1]{fontenc}
\usepackage{hyperref}
\usepackage{graphicx}
\usepackage{listings}
\usepackage{physics}
\usepackage[normalem]{ulem}

\usepackage{amsthm}
\usepackage{amsmath}
\usepackage{amssymb}
\usepackage{mathtools}
\usepackage[colorinlistoftodos]{todonotes}
\usepackage{tikz}
\usetikzlibrary{matrix,fit,calc,positioning}

\newtheorem{theorem}{Theorem}
\newtheorem{corollary}[theorem]{Corollary}

\newcommand{\tit}{Minimal qubit representations of Hamiltonians via conserved charges}

\newcommand{\CliffC}{\mathcal{C}}
\newcommand{\RR}{\mathcal{RR}}
\newcommand{\CR}{\mathcal{CR}}

\begin{document}

\preprint{APS/123-QED}

\title{
 \tit
}

\author{Lane G. Gunderman}
\thanks{Equal contribution author}
\email{lanegunderman@gmail.com}
\affiliation{Independent scholar}
\author{Andrew Jena}
\thanks{Equal contribution author}
\email{ajjena@uwaterloo.ca}
\affiliation{
Institute for Quantum Computing, University of Waterloo, Waterloo, ON N2L 3G1, Canada
}
\affiliation{Department of Combinatorics and Optimization, University of Waterloo, Waterloo, Ontario, N2L 3G1, Canada}
\author{Luca Dellantonio}
\email{l.dellantonio@exeter.ac.uk}
\affiliation{Department of Physics and Astronomy, University of Exeter, Stocker Road, Exeter EX4 4QL, United Kingdom}

\date{\today}

\begin{abstract}
In the last years, we have been witnessing a tremendous push to demonstrate that quantum computers can solve classically intractable problems. This effort, initially focused on the hardware, progressively included the simplification of the models to be simulated. We consider Hamiltonians written in terms of Pauli operators and systematically cut all qubits that are not essential to simulate the system. Our approach is universally applicable and lowers the complexity by first ensuring that the largest possible portion of the Hilbert space becomes irrelevant, and then by finding and exploiting all conserved charges of the system, i.e., symmetries that can be expressed as Pauli operators. Remarkably, both processes are classically efficient and optimal. To showcase our algorithm, we simplify chemical molecules, lattice gauge theories, the Hubbard and the Kitaev models.
\end{abstract}

\maketitle


%
\section{
Introduction
}\label{sec:intro}
Today, we are witnessing tremendous efforts for enhancing the gate fidelities and incrementing the number of qubits in all types of quantum hardware. However, reaching the thresholds that are required for fault-tolerant quantum computation \cite{Aharonov2008,Kitaev2003} is still very challenging. As such, it is of paramount importance to lower the complexity of the models to be simulated as much as possible, such that in the near term more and more practical applications can enjoy a quantum advantage. For instance, approaches such as the variational quantum eigensolver \cite{peruzzo2014variational,chan2023hybrid,Ferguson2021,verteletskyi2020measurement,gokhale2019minimizing,kandala2017hardware,mazzola2023quantum} rely on understanding and simplifying the Hamiltonian to design tailored and minimal resources for the experiment (see, e.g., Refs.~\cite{mishmash2023hierarchical,Choi2023fluidfermionic,richerme2023quantum,sparrow2018simulating,Wang2020efficient} and Refs.~\cite{paulson2021,Haase2021resourceefficient,khaneja2000cartan,Mathis2020,Mazzola2021gauge,funcke2023exploring,celi2020emerging,meth2023simulating} for chemistry and lattice gauge theories, respectively). But the search for simplification is much wider, including (besides countless others) quantum machine learning \cite{torlai2018neural,schuld2014quest,jeswal2019recent,beer2020training}, optimization tasks \cite{cain2023quantum,Moll2018quantum,MiguelRamiro2023optimizedquantum,ramiro2022collective}, and tensor network techniques \cite{tagliacozzo2014tensor,singh2011tensor,orus2019tensor,cirac2021matrix,Mortier2022,Rader2018,Corboz2018,Schuch2013,Delcamp2021tensornetwork,Vanderstraeten2017}.

Within the quantum simulation and computation communities, however, an approach that is (a) universally applicable and (b) classically efficient is still missing. In this work, we consider a Hamiltonian $\mathcal{H}$ expressed as a sum of Pauli operators 
\footnote{
Notice that any $d$-sparse Hamiltonian can be efficiently embedded in an invariant $SU(2)$ subspace, see \cite{Low2019hamiltonian}
}
and determine a Clifford transformation $\mathcal{C}$ such that $\mathcal{C}\mathcal{H}\mathcal{C}^{\dagger}$ is block diagonal with the blocks being as small as possible. In other words, we provide an algorithm that eliminates \textit{all} superfluous qubits by determining redundancies and conserved charges of the model. Remarkably, these are found classically, efficiently, and without any prior knowledge. Besides the qubit reduction our work allows for parallelization of quantum simulations on both classical and quantum platforms, and individual study of independent subsectors of the Hamiltonian.

\section{
Main Results
}\label{sec:main_res}
\begin{figure}
\centering
\includegraphics[width=\columnwidth]{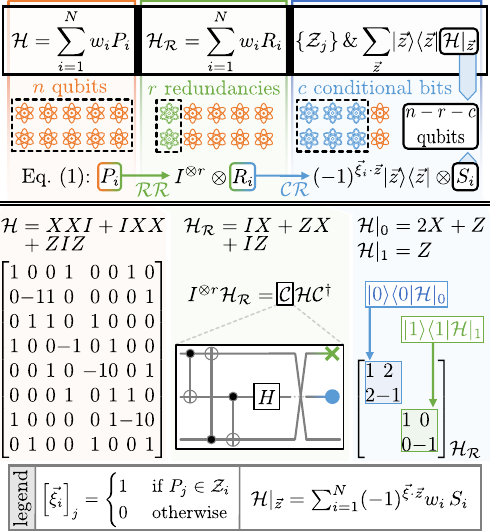}
\caption{
\textit{(top)}: Underlying idea. From left to right, an input $\mathcal{H}$ is given, acting on $n$ qubits. $\RR$ first determines a minimal representation $\mathcal{H}_{\mathcal{R}}$, equivalent to $\mathcal{H}$ but requiring $n-r$ qubits. Afterwards, $\CR$ finds all conserved charges $\{ \mathcal{Z}_j \}_{j=1}^c$ of $\mathcal{H}_{\mathcal{R}}$ (equiv., $\mathcal{H}$). Employing their eigenvectors $\ket{\vec{z}}$, the problem is further simplified to the $n-r-c$ qubits $\mathcal{H}|_{\vec{z}}$. The whole process is well represented by the operators $P_i$, $R_i$ and $S_i$ in the Hamiltonians $\mathcal{H}$, $\mathcal{H}_{\mathcal{R}}$ and $\mathcal{H}|_{\vec{z}}$, respectively. \textit{(bottom)}: Example from a modified $J_1-J_2$ model with $XX$ ($ZZ$) interactions between (next) neighbouring spins \cite{majumdar1969next,majumdar1969next2} and $J_1 = J_2 =1$. Via the Clifford circuit $\mathcal{C}$ consisting of $H$, $CNOT$, and $SWAP$ gates, the input matrix is both reduced dimensionally and made block-diagonal, with blocks describing $n-r-c=1$ qubit. Qubits removed via $\RR$ and $\CR$ are indicated in $\mathcal{C}$ with the green cross and blue circle, respectively.
}
\label{fig:scheme}
\end{figure}

As shown in Fig.~\ref{fig:scheme}\textit{(top)}, we consider an $n$-qubit system whose dynamics is governed by the Hamiltonian
\begin{equation}\label{eq:Hamiltonian_input}
    \mathcal{H}
    =
    \sum_{i=1}^{N} w_i P_i, 
    \text{ where } 
    P_i 
    \overset{\text{e.g.}}{=} 
    \underbrace{I \otimes X \otimes Y \dots Z \otimes I \otimes X}_{n \text{ qubits}}
    .
\end{equation}
Here, $\lbrace \omega_i \rbrace_{i=1}^{N}$ are real numbers and $\{ I,X,Y,Z \}$ the identity and the three Pauli operators, respectively. For clarity we focus on Hamiltonians expressed as qubit Pauli operators. The generalization to qudits can be done following similar steps as for the qubits, see the Appendix for details. To simplify $\mathcal{H}$ into the desired form, the main tool we employ is the Clifford group, under the form of a circuit $\CliffC$ consisting of Hadamard $H$, phase $S$, and the entangling $CNOT$ gates (for an example, see bottom panel in the figure). 

The advantages from the Clifford group are two-fold. First, by definition \cite{gottesman1998heisenberg,aaronson2004improved}, it is the most general framework to perform a basis change of the Hamiltonian $\mathcal{H} \xrightarrow[]{} \CliffC \mathcal{H} \CliffC^\dagger$ while retaining the same form as in Eq.~\eqref{eq:Hamiltonian_input}. For each $P_i$,
$\CliffC P_i \CliffC^\dagger$ is a Pauli operator, and conversely each unitary transforming Pauli operators into Pauli operators can be described as a Clifford circuit. 

The second advantage is the efficiency. Simulating Clifford circuits on \textit{classical} computers requires polynomial resources with respect to the number of qubits \cite{gottesman1998heisenberg,aaronson2004improved}. For the same reasons,  for $n$ qubits and $N$ Pauli operators our algorithm has execution times that scale as $O(n^{2} N)$.
Furthermore, our method is equally applicable to simulations on classical or quantum devices. Indeed, we both provide (see Sec.~\ref{sec:algo}) the Hamiltonian in its simplified form \textit{and} the circuit $\CliffC$ that can be readily implemented on every quantum setup.
%

Fig.~\ref{fig:scheme} illustrates our approach, with the top panel displaying the algorithmic subroutines on which it is based. Their high-level explanation is outlined in this section, alongside the example in the bottom panel.
The main theoretical proofs demonstrating the feasibility \textit{and} the optimality of our formalism are given in Sec.~\ref{sec:algo}, with the details outlined in the Appendix. Finally, in Secs.~\ref{sec:res} and \ref{sec:conclusions} we apply our algorithm to several physical models and provide concluding remarks, respectively.

We begin by reducing the number of qubits required for exactly representing the full system dynamics. This is the ``Redundancy Removal'' ({$\RR$}) in Fig.~\ref{fig:scheme} (green box), which effectively eliminates the trivial portion of the $n$-qubits Hilbert space. With the scope of lowering the complexity of a considered problem, similar ideas have been always utilized in physics. For instance, in the case of Fermionic systems with a fixed number of excitations, a minimal qubit representation was found in Ref. \cite{harrison2022reducing}. However, this and other works \cite{mishmash2023hierarchical,khaneja2000cartan,gorshkov2007universal} either consider specific (classes of) Hamiltonians, instead of an arbitrary $\mathcal{H}$ as in Eq.~\eqref{eq:Hamiltonian_input}, or were not ensured to be (classically) efficient nor optimal. Our approach is completely general, efficient \textit{and} optimal (see Theorem~\ref{thm:theorem1}), implying that we deterministically find the smallest possible subspace of the system Hilbert space to faithfully describe the whole dynamics.

In more details, {$\RR$} works as follows \footnote{
Note that a subroutine similar to {$\RR$} was previously investigated in Ref. \cite{gunderman2023transforming}.
}. As schematically represented in Fig.~\ref{fig:scheme}, our algorithm determines a Clifford circuit $\CliffC$ such that $\CliffC P_i \CliffC^\dagger = \left( \bigotimes_{k=1}^{r} I \right)\otimes R_i$ for all $i=1,\dots,N$.
Consequently, the first $r$ qubits become negligible and
the system Hamiltonian in Eq.~\eqref{eq:Hamiltonian_input} can be represented by
$\mathcal{H} \xrightarrow[]{\RR} \mathcal{H}_{\mathcal{R}} = \sum_{i=1}^{N} \omega_i R_i$.

An example of the $\RR$ subroutine is given in Fig.~\ref{fig:scheme}\textit{(bottom)}. We consider $\mathcal{H} = XXI + IXX + ZIZ$ on $n=3$ qubits, and employ our protocol to determine the Clifford circuit $\mathcal{C}$ in the figure. None of the terms in the resulting $\CliffC \mathcal{H} \CliffC^\dagger = IIX + IZX + IIZ$ acts on the first $r=1$ qubit, such that the whole system dynamics is described via the operator $\mathcal{H}_{\mathcal{R}}$ defined on a reduced Hilbert space of $n-r = 2$ qubits.

Our second reduction, called ``Conditional Removal'' ($\CR$), leverages conserved charges. These, indicated with $\{ \mathcal{Z}_j \}_{j=1}^{c}$ in Fig.~\ref{fig:scheme}, are the $c$ Pauli symmetries of the Hamiltonian $\mathcal{H}_{\mathcal{R}}$, meaning that for each $j=1,\dots,c$, $[\mathcal{Z}_j,\mathcal{H}_{\mathcal{R}}] = 0$ and $\mathcal{Z}_j$ is a tensor product of $n-r$ Pauli operators. Importantly, our approach ensures that these conserved charges fulfil two properties. First, $\CR$ finds \textit{all} of them, meaning that \textit{any} Pauli symmetry of $\mathcal{H}_{\mathcal{R}}$ can be written as a linear combination of the $\{ \mathcal{Z}_j \}_{j=1}^{c}$. Second, \sout{that} we can express the conserved charges in the practical form
%
\begin{equation}\label{eq:conserved_charges}
    \mathcal{Z}_j = 
    \underbrace{
    I \otimes I \dots \otimes 
    \overbrace{
    Z
    }
    ^{j^{\rm th}}
    \otimes \dots \otimes I
    }
    _{c \text{ qubits}}
    \bigotimes_{i=1}^{n-r-c} I
    ,
\end{equation}
i.e., in terms of $c$ diagonal, independent generators with a single $Z$ operator each. This form for the conserved charges can be chosen since they must commute with each other
and as such they can be efficiently diagonalized \cite{jena2019pauli,shlosberg2021adaptive}. 

Expressing the conserved charges $\mathcal{Z}_j$ as in Eq.~\eqref{eq:conserved_charges} is not necessary. One can better understand the symmetries of
$\mathcal{H}$ in Eq.~\eqref{eq:Hamiltonian_input}, indicated with $\{ \tilde{\mathcal{Z}}_j \}_{j=1}^c$, by reverting the basis change characterized by $\CliffC$ to find that $[\tilde{\mathcal{Z}}_j,\mathcal{H}] = 0$ for $\tilde{\mathcal{Z}}_j \equiv \CliffC^\dagger ((\bigotimes_{i=1}^{r}I )\otimes \mathcal{Z}_j) \CliffC$.
Furthermore, an interesting question 
is whether it is feasible to lower the complexity of simulating $\mathcal{H}_{\mathcal{R}}$ by changing the form of the conserved charges (e.g., concentrating the system entanglement in the $c$ qubits onto which they act).

\begin{figure}
\centering
\includegraphics[width=\columnwidth]{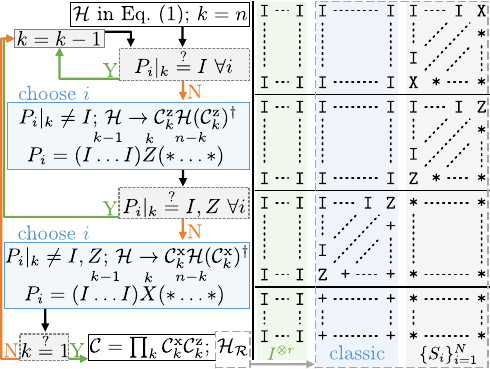}
\caption{
\textit{(left)}: Our algorithm. From the input $\mathcal{H}$ and on the $k$-th qubit ($k=n,n-1,\ldots,1$), in the first [second] step we iteratively check (gray boxes with dashed lines) whether all $P_i$ acting on the $k$-th qubits are all $I$ [$I$ or $Z$]. If they are, we successfully identified a redundant [conditional] qubit and proceed. If they are not, we pick a $P_i$ such that $ P_i|_k \neq I$ [$P_i|_k \neq I,Z$] and build (lightblue boxes) the $\mathcal{C}_{k}^{\rm z}$ [$\mathcal{C}_{k}^{\rm x}$] that ensures $ P_i|_j = I$ for all $j=1,\ldots,k-1$, and $ P_i|_k = Z$ [$P_i|_k = X$]. 
This process ensures that $\mathcal{H}_{\mathcal{R}}$ is as in the \textit{(right)} panel, i.e., the first $r$ qubits are only acted upon by $I$ and the next $c$ by $I$ or $Z$. The operators $S_i$ are determined from the last $n-r-c$ column of the tableau, $+$ and $*$ are used to indicate $I$ or $Z$, and any Pauli operator, respectively. The coefficients (omitted for clarity) are tracked with their associated Pauli operator.
}
\label{fig:algo}
\end{figure}

In the context of this work and the $\CR$ subroutine specifically, 
we express the conserved charges as in Eq.~\eqref{eq:conserved_charges} for clarity and functionality. In fact, in this form the computational basis $\ket{\vec{z}}$ with $\vec{z} = \{ 0,1 \}^{c}$ (e.g., $\vec{z} = 0111001$ for $c=7$) contains all eigenvectors of the conserved charges $\{ \mathcal{Z}_j \}_{j=1}^{c}$, and therefore of the subset of corresponding $c$ qubits of the $\{ R_i \}_{i=1}^{N}$ operators within $\mathcal{H}_{\mathcal{R}}$ (see Fig.~\ref{fig:scheme}). Even better, we can determine the associated charge eigenvalues $(-1)^{\vec{z}\cdot \vec{\zeta}_i}$ and consequently divide $\mathcal{H}_{\mathcal{R}}$ into $2^c$ separated, non-interacting subsectors, each governed by its own Hamiltonian $\mathcal{H}\vert_{\vec{z}}$
\begin{subequations} \label{eq:Hamiltonian_out}
  \begin{align}
     & \mathcal{H}_{\mathcal{R}}
    =
    \sum_{\vec{z}}
    \ketbra{\vec{z}}{\vec{z}}
    \otimes
    \overbrace{
    \sum_{i=1}^{N}
    (-1)^{\vec{z}\cdot\vec{\zeta}_i}
    \omega_i
    S_i    
    }^{
    \mathcal{H}\vert_{\vec{z}}
    }    ,\label{eq:Hamiltonian_out_H} \\
     & 
     \left[\vec{\zeta}_{i}\right]_j = 1 
     \text{ if }
     \mathcal{Z}_j \in R_i,
     \text{ }0\text{ otherwise}
     . \label{eq:Hamiltonian_out_S_theta}
  \end{align}
\end{subequations}
Here, ``$\cdot$'' is the dot product modulo $2$, $\left[\vec{\zeta}_{i}\right]_j$ is the $j$-th component of $\vec{\zeta}_i$ and $\mathcal{Z}_j \in R_i$ indicates that $R_i$ has nontrivial support from $\mathcal{Z}_j$. The $S_i$ are the last $n-r-c$ Pauli terms within $R_i$ [see Fig.~\ref{fig:algo}\textit{(right)}].

The physical intuition behind Eqs.~\eqref{eq:Hamiltonian_out} is the following. After $\RR$ and the removal of $r$ redundant qubits, the trivial dynamics has been fully eliminated. Yet, via $\CR$, it is still possible to efficiently analyze $c$ out of the remaining $n-r$ qubits via the conserved charges $\{ \mathcal{Z}_j \}_{j=1}^{c}$. Practically, this is done by collapsing the system dynamics onto their eigenvectors $\ket{\vec{z}}$ [hence the projectors in Eq.~\eqref{eq:Hamiltonian_out_H}] and via the corresponding sub-Hamiltonians $\mathcal{H}|_{\vec{z}}$. These are determined from $\mathcal{H}_{\mathcal{R}}$ by writing its constituents $R_i$ as $R_i = \sum_{\vec{z}} (-1)^{\vec{z} \cdot \vec{\zeta}_i} \ketbra{\vec{z}} \otimes S_i$. 

The family of sub-hamiltonians $\mathcal{H}|_{\vec{z}}$ acting on $n-r-c$ qubits fully describe the whole system. The union of their eigenvalues (eigenvectors) are the eigenvalues (eigenvectors -- up to the basis change $\CliffC$) of the original Hamiltonian $\mathcal{H}$. And the time evolution of $\ket{\vec{z}}\otimes \ket{\psi_{\rm in}}$ can be written as $\exp(
-i \mathcal{H}_{\mathcal{R}} t
)
(
\ket{\vec{z}}
\otimes 
\ket{\psi_{\rm in}}
)
= 
\ket{\vec{z}}
\otimes 
\exp(
-i \mathcal{H}|_{\vec{z}} t
)
\ket{\psi_{\rm in}}$ for any time $t$ and states $\ket{\vec{z}}$, $\ket{\psi_{\rm in}}$ \footnote{
Up to a known phase that here is omitted for clarity. For the same reason, we set $\hbar=1$.
}.

Going back to the example in Fig.~\ref{fig:scheme}, above we employed $\RR$ to describe the system via the $n-r = 2$ qubits Hamiltonian $\mathcal{H}\xrightarrow[]{\RR}\mathcal{H}_{\mathcal{R}} = IX + ZX + IZ$. As shown in the figure, our algorithm ensures (via the Clifford circuit $\CliffC$) that the matrix form of $\mathcal{H}_{\mathcal{R}}$ is block diagonal. These blocks, each of size $2^{n-r-c} = 2$, encode the $n-r-c = 1$ qubit Hamiltonians $\mathcal{H}|_{\vec{z}}$ corresponding to the eigenvectors $\ket{\vec{z}} = \ket{0}$ and $\ket{\vec{z}} = \ket{1}$ of the single charge $\mathcal{Z}_1 = Z\otimes I$. Therefore, the $\CR$ subroutines permits rewriting $
\mathcal{H}_{\mathcal{R}}
\xrightarrow[]{\CR} 
\ketbra{0}\left[ X + X + Z \right]
+
\ketbra{1}\left[ X - X + Z \right]
$, where the sign of each term in the square parentheses depends on the charge eigenvalue of $\ket{\vec{z}}$ and whether $\mathcal{Z}_1 \in R_i$.

Concluding, Eqs.~\eqref{eq:Hamiltonian_out} are the mathematical tools that allows for simplifying qubit-based 
physical models.
They tell us that when conserved charges are present (and our method finds all of them), simulating the system as a whole is not necessary. We can restrict ourselves to smaller subsectors that are generally much easier to be dealt with. Furthermore, knowing the conserved charges allows the identification of the subsectors describing the dynamics of interest. For instance, only the case of zero total electrical charge is usually investigated in $\mathbb{Z}_2$ models of lattice gauge theory (see Sec.~\ref{sec:res}), making most of the Hilbert space irrelevant. Our approach allows determining the Hamiltonian describing the corresponding dynamics, resulting in a significant reduction of the simulation complexity.

Finally, even in the scenario in which all $2^c$ subsectors must be studied, our approach allows for complexity reduction and/or parallelization of quantum processes, as each of the $\mathcal{H}|_{\vec{z}}$ Hamiltonians can be individually addressed. In the classical case, diagonalization algorithms require runtimes $O(2^{\alpha n})$, with $2 \leq \alpha \leq 3$ \cite{banks2022pseudospectral,alman2021refined,armentano2018stable}. Therefore, (without) studying the $2^c$ Hamiltonians $\mathcal{H}|_{\vec{z}}$ in parallel, our method allows reducing the total time from $2^{\alpha n}$ to $2^{\alpha(n-r-c)}$ ($2^c \cdot 2^{\alpha(n-r-c)}$), or by a factor of $2^{\alpha(r + c)}$ ($2^{\alpha r + (\alpha-1)c}$), which can be drastic. In the quantum case, on the other hand, Eqs.~\eqref{eq:Hamiltonian_out} imply that all eigenstates of $\mathcal{H}_{\mathcal{R}}$ are such that the first $c$ qubits are in the computational basis. By taking the variational quantum eigensolver \cite{peruzzo2014variational,chan2023hybrid,Ferguson2021,verteletskyi2020measurement,gokhale2019minimizing,kandala2017hardware} as an example, this means that these qubits only require local gates to reach the groundstate, with possible reductions in circuit complexity and depth. 
%


%
\section{
Algorithm
}\label{sec:algo}
Our algorithm is described in Fig.~\ref{fig:algo}. We represent $\mathcal{H}$ in Eq.~\eqref{eq:Hamiltonian_input} as an $N \times n$ tableau \footnote{
As better explained in Sec.~\ref{sec:algo} and in App.~\ref{app:opt_proof}, our algorithm does not work with all $N$ elements of the input Hamiltonian $\mathcal{H}$. Instead, it employs a generating subset that can be much smaller in practice, such that the efficiency is enhanced.
} with entries $I$, $X$, $Y$ and $Z$ [see Fig.~\ref{fig:algo}\textit{(right)}], where the $i$-th row corresponds to $P_i$ and the $k$-th column to the qubit. Therefore, the $(i,k)$ element $P_{i}|_{k}$ is the Pauli operator acting on the $k$-th qubit of $P_i$. 

With $\mathcal{H}$ written as a tableau, one can identify the set of standard operations corresponding to Clifford gates (see App.~\ref{app:opt_proof} and Ref.~\cite{aaronson2004improved}). These, in turn, are employed in the circuits $\CliffC_{k}^{\rm z}$ and $\CliffC_{k}^{\rm x}$ used in the $k$-th iteration of the algorithm to build $\CliffC = \prod_{k} \CliffC_{k}^{\rm x} \CliffC_{k}^{\rm z}$ [see Fig.~\ref{fig:algo}\textit{(left)}]. $\CliffC$ is the Clifford circuit required by the $\RR$ subroutine to determine $\mathcal{H}_{\mathcal{R}}$.
This is deduced from the two blue boxes in the figure, explaining that $\CliffC$ is constructed to ensure only identities act on the first $r$ qubits, while $X$ and $Y$ are collected in the last $n-c-r$, leaving the $c$ in the middle with only $I$ or $Z$
\footnote{
Additional $SWAP$ operations may be required. For clarity, we did not explicitly include those in the algorithm.
}.

Our algorithm in Fig.~\ref{fig:algo}\textit{(left)} always ensures $\mathcal{H}_{\mathcal{R}}$ is block diagonal,
as its first $c$ qubits are acted upon by diagonal operators. To find the form of the block $\mathcal{H}|_{\vec{z}}$, the $\CR$ subroutine first determines the $\{ S_i \}_{i=1}^{N}$ from the last $n-r-c$ column of the tableau in Fig.~\ref{fig:algo}\textit{(right)}. Then, depending on the chosen $\vec{z}$, the corresponding signs are calculated from the $c$ columns with only $I$ and $Z$.

With more details presented in App.~\ref{app:opt_proof}, we demonstrate here the optimality of our algorithm. I.e., both the reduction of $r$ redundant qubits and the number $c$ of conserved charges are the maximal ones. This is stated by the following 
\begin{theorem}
\label{thm:theorem1}
Let $\mathcal{H}$ be as in Eq.~\eqref{eq:Hamiltonian_input} and $M$ the matrix of the commutation coefficients for any generating subset of the $\{ P_i \}_{i=1}^{N}$. Then, for any choice $0\leq c \leq \dim(M)-\rank(M)$, $\mathcal{H}$ can be simulated using $\frac{1}{2}\rank(M)+[\dim(M)-\rank(M)-c]$ qubits and a maximum of $2^c$ subproblems.
\end{theorem}
A generating subset of the $\{ P_i \}_{i=1}^{N}$ is a product-wise independent collection of elements such that any of the $P_i$ can be expressed as a product of those. The element $M_{ij}$ of $M$ is one if $[P_i,P_j] \neq 0$, zero otherwise, and the $2^c$ subproblems are (e.g.) the $\mathcal{H}|_{\vec{z}}$ defined above, that generally depend on the the chosen generating subset (albeit their size does not). $\dim(M)$ and $\rank(M)$ are the dimension and the rank (computed over the field of integers modulo $2$) of $M$, respectively.

The detailed proofs of Theorem~\ref{thm:theorem1} and the optimality of the algorithm in Fig.~\ref{fig:algo} are left to App.~\ref{app:opt_proof}. Intuitively, the first relies on symplectic linear algebraic subspaces, while the latter follows from the properties of the resulting $\mathcal{H}|_{\vec{z}}$. Indeed, they saturate the bound $c = \dim(M)-\rank(M)$ by construction, i.e., their size is the minimal possible one that is achieved via optimal $\RR$ and $\CR$ processes. 

\section{
Results
}\label{sec:res}
\begin{figure}
\centering
\includegraphics[width=\columnwidth]{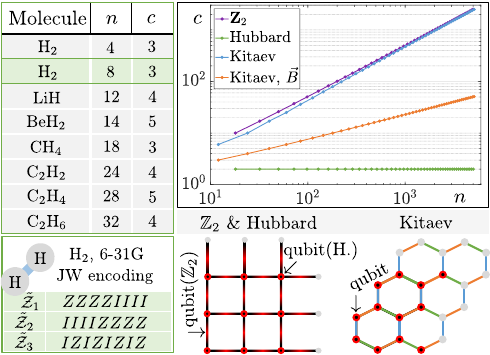}
\caption{
Values of $c$ for \textit{(left)} chemistry Hamiltonians in Refs.~\cite{mcclean2020openfermion,chem_repo} and \textit{(right)} the $\mathbb{Z}_2$ \cite{Ferguson2021,chan2023hybrid,Haase2021resourceefficient,Horn1979}, Hubbard \cite{stanisic2022observing} and Kitaev (with and without magnetic field $\vec{B}$) \cite{bespalova2021quantum,KITAEV20062} models -- additional information in App.~\ref{app:phys_mod}. Below the table, we write the conserved charges for $\text{H}_2$ (sketch) in the $6-31G$ basis and JW encoding. For the other molecules we employ the $STO-3G$ basis.
Below the plot, we draw the smallest instances of the square and hexagonal lattices employed. Grey vertices indicate periodic boundary conditions. 
}
\label{fig:res}
\end{figure}

In Fig.~\ref{fig:res} we present the results from our algorithm applied to several physical models. In the left, we consider chemistry Hamiltonians (equivalently in the Jordan-Wigner (JW), Bravyi-Kitaev (BK),  or Parity encoding \cite{cao2019quantum,stanisic2022observing,chem_repo}) and empirically show that, albeit molecule dependent, the number of conserved charges scales as a constant with respect to the molecule size. This is unsurprising, as molecules generally have limited symmetries associated to conserved charges \cite{lowe2011quantum}. In previous works \cite{richerme2023quantum,kandala2017hardware,Hempel2018quantum,obrien2022purificationbased,Elfving_2021}, some were employed to simulate small molecules on quantum computers. Specifically, finding the reduced Hamiltonians of H$_2$ \cite{richerme2023quantum} and LiH \cite{kandala2017hardware} required significant efforts. Not only is our algorithm capable of doing the same for any molecule; it also ensures all conserved charges are found. For instance, the ones of $\text{H}_2$ in the JW encoding are reported at the bottom left of the figure.

In the right panels of the figure we investigate the $\mathbb{Z}_2$ \cite{Ferguson2021,chan2023hybrid,Haase2021resourceefficient} (violet), Hubbard \cite{stanisic2022observing} (green), Kitaev (blue) and Kitaev with magnetic field \cite{bespalova2021quantum,KITAEV20062} (orange) models. The smallest lattices we employed are at the bottom, where we indicate whether qubits lie on vertices (Hubbard and Kitaev) or edges ($\mathbb{Z}_2$). Larger instances of the models are obtained by increasing the number of plaquettes/hexagons in both dimensions at the same time (for more details, see App.~\ref{app:phys_mod}).

From the plot we deduce that the Hubbard model has a constant number $c=2$ of conserved charges, corresponding to global properties (see App.~\ref{app:phys_mod}). For $\mathbb{Z}_2$, describing (discretized) cavity electromechanics on a torus, our algorithm efficiently implements all $c=n/2+1$ Gauss laws and determines the Hamiltonians with any configuration of static charges \cite{paulson2021}. For the Kitaev honeycomb lattice \cite{bespalova2021quantum}, when a magnetic field $\vec{B}$ is present, we still find $c=(\sqrt{2n+1}+1)/2$ conserved charges that can be exploited to both better understand the model and significantly enhance ongoing classical and quantum simulations of the system \cite{bespalova2021quantum}. Without $\vec{B}$, an analytical solution of the Kitaev honeycomb lattice is known \cite{KITAEV20062}, however our algorithm yields $n-c \neq cnst$. This is because not all the symmetries correspond to conserved charges. As better discussed in the conclusions, there is a clear road map to extend our method to identify and exploit more symmetries and thus further simplify input models.

\section{
Conclusions
}\label{sec:conclusions}
In this work, we designed an algorithm based on similar techniques as in the Gottesman-Knill theorem \cite{gottesman1998heisenberg,aaronson2004improved} to find the Clifford equivalent Hamiltonian that requires the smallest number of qubits to simulate the full system dynamics.
Our framework is ideal for further simplifications (see below), and allowed us to develop an algorithm that is classically efficient and optimal. Here, optimal means that it deterministically removes all redundant qubits ($\RR$ subroutine) and exploits all conserved charges ($\CR$ subroutine). We showcased our method with chemical molecules, $\mathbb{Z}_2$ cavity electrodynamics, the Hubbard and the Kitaev (with and without a magnetic field) models. In all cases, we were able to find symmetries that can lower the complexity of classical and quantum simulations. The number of qubits that can be eliminated vary from a constant (chemistry molecules and the Hubbard model) to a fixed ratio of the total. For the Kitaev model with external magnetic field $\vec{B}$ we find $\mathcal{O}(\sqrt{n/2})$ symmetries, which becomes $\mathcal{O}(n/2)$ for both the $\mathbb{Z}_2$ and the Kitaev without $\vec{B}$ models.

From Fig.~\ref{fig:res}, we deduce that the numbers of conserved charges in the Hubbard and Kitaev (without $\vec{B}$) models are less than the known symmetries of these systems. In some cases (e.g., Ref.~\cite{KITAEV20062}), it is even possible to find an analytical solution. Generally, this requires significant efforts that we want to eliminate via our automatized framework. To do so, we will extend this work to include, besides others, (1) composite qudit Hamiltonians \footnote{The prime and prime power cases are handled in the Appendix, while a non-Clifford, and not per se optimal, solution has been shown in \cite{sarkar2023qudit}}, (2) Clifford symmetries
\footnote{
Symmetries with respect to elements of the Clifford group.
}, and
(3) separability criteria. 
However, our results can already be used to shed light on open questions. By finding the numbers and types of conserved charges in chemical molecules [see Fig.~\ref{fig:res}\textit{(left)}], it can be better understood whether the technique in Ref.~\cite{huggins2022unbiasing} scales favourably in the molecule size \cite{mazzola2022exponential}. 

To facilitate the advent of useful quantum computation, the theory must provide simplified models that alleviate the requirements on available and near-term hardware.
The Clifford group holds the potential to allow for this simplification, yet 
there are exponentially many 
$n$-qubits Clifford unitaries to choose from. This work is a first step towards finding the ones that do provide an advantage. This advantage stems from the ability of efficiently determining all conserved charges and the associated reduction of qubits required for the simulation. 

\section*{Additional informations}
The algorithm described in this work can be found at the github repository in Ref.~\cite{github_repo}. After we wrote this manuscript, we became aware of Ref.~\cite{bravyi2017tapering}, where a similar procedure for tapering off qubits is outlined. Compared to \cite{bravyi2017tapering}, we provided a proof of optimality and demonstrated the applicability of our algorithm to different physical models.

\section*{Acknowledgments}
AJ and LD are grateful to Michele Mosca for the constant support. We thank Wolfgang D{\"u}r and Vincent Elfving for helpful discussions.

\section*{Funding}
This work was supported by NTT Research and the EPSRC quantum career development grant EP/W028301/1

\bibliographystyle{unsrt}
\phantomsection  
\renewcommand*{\bibname}{References}

\clearpage
\appendix
\section*{Appendix}
Below, we provide additional details to ``{\tit}''. Firt, we demonstrate the optimality of Theorem \ref{thm:theorem1} and our algorithm. Second, we describe in more details the physical models considered in Sec.~\ref{sec:main_res} for the numerical results. Finally, we give the pseudocode of our algorithm and a table collecting additional numerical results.
\section{Proof of Optimality}
\label{app:opt_proof}
The Pauli group for a single qubit is generated by the $X$ and $Z$ Pauli operations. These are given by:
\begin{equation}
    X=\begin{bmatrix}
    0 & 1\\
    1 & 0
    \end{bmatrix},\quad Z=\begin{bmatrix}
    1 & 0\\
    0 & -1
    \end{bmatrix},
\end{equation}
with computational basis states $[1\ 0]$ and $[0\ 1]$ and a general qubit being a normalized vector in $\mathbb{C}^2$. The last non-trivial member of the group is called $Y$ and is proportional to $X\cdot Z$. In the case of multiple qubits, we take tensor products of the four single qubit Pauli group members $\{I,X,Y,Z\}$. A useful alternative picture for Pauli operators is the symplectic representation \cite{nielsen2002quantum,lidar2013quantum,gunderman2020local}. In this representation a Pauli operator on $n$ qubits is written as a $2n$ entried binary vector with a single leading scalar. For our work we define this as:
\begin{equation}
    w_iP_i:= w_i \cdot X^{\vec{a_i}}\cdot Z^{\vec{b_i}}\mapsto (w_i\ |\ \vec{a_i}\ |\ \vec{b_i}),
\end{equation}
with $w_i$ being a coefficient for the Pauli operator $P_i$. For simplicity we will often drop the coefficient entry since it does not impact the commutation structure, and may include this coefficient again at the end. If we have multiple Pauli operators we are representing in the symplectic representation, we may use them as rows in a matrix. Lastly, we recall that the Clifford group is the set of gates which upon conjugating a Pauli operator still produces a Pauli operator. The Clifford group is classically easy to simulate and is generally considered a rather early primitive operation for quantum computers. These Clifford operations then map symplectic vectors representing Pauli operators to other symplectic vectors representing Pauli operators. In this appendix we use the term \textit{register} usually as the more general term for qubit \cite{watrous2018theory}.


Here we show how to minimally represent the commutation relations for a collection of Pauli operators. The resulting Pauli operators will be related to the original ones through a Clifford circuit, which is classically easy to perform. In this way we may reduce the number of qubits needed to perform a circuit equivalent to the initially given Pauli one. This result is largely the same as \cite{gunderman2023transforming}, however, these results are taken further here. While only Clifford operations are permitted, we may use linear row addition in the matrix to determine the number of generators needed to generate the full set of Pauli operators.

Let $\mathcal{P}$ be the given collection of Pauli operators. The elements of $\mathcal{P}$ can be generated by compositions of generators, $\mathcal{G}$, of the smallest subgroup of Pauli operators which contains $\mathcal{P}$. Then define the commutation matrix $M(\mathcal{G})$ as:
\begin{equation}
    M(\mathcal{G})_{ij}=\mathcal{G}_i\odot \mathcal{G}_j,
\end{equation}
with $\mathcal{G}_i\odot \mathcal{G}_j=0$ if they commute, and $1$ otherwise. Formally this is the symplectic product of the symplectic representations of the generators \cite{ketkar2006nonbinary,gunderman2020local}.

The commutation matrix itself is not invariant under the selection of the generators in $\mathcal{G}$. Replacing generator $\mathcal{G}_1$ with $\mathcal{G}_1\circ \mathcal{G}_2$ is equivalent to adding the row corresponding to $\mathcal{G}_2$ to the row corresponding to $\mathcal{G}_1$, and likewise adding the same column. This forms a symmetric Gaussian elimination technique and so the \textit{rank} of $M(\mathcal{G})$ is unaltered under the selection of generators \footnote{Formally the rank is computed for the matrix over the field of integers modulo $2$}.

Using this, there is a set of generators $\mathcal{D}$ such that:
\begin{equation}
    M(\mathcal{D})=\left(\bigoplus_{i=0}^{dim(M)-rank(M)} [0]\right)\bigoplus \left(\bigoplus_{i=0}^{\frac{1}{2}rank(M)} \begin{bmatrix}
    0 & 1\\
    1 & 0
    \end{bmatrix}\right).
\end{equation}
This can be satisfied by generators $Z_i$ for the first $dim(M)-rank(M)$ registers, and anti-commuting pairs $\{X_i,Z_i\}$ for the remaining registers. This means that to represent the commutation relations for $\mathcal{P}$, only $dim(M(\mathcal{G}))-\frac{1}{2}rank(M(\mathcal{G}))$ registers are needed. It is not possible to use fewer registers than this and still be Clifford equivalent to the original collection $\mathcal{P}$. This means that if the original collection $\mathcal{P}$ operated on $t>dim(M(\mathcal{G}))-\frac{1}{2}rank(M(\mathcal{G}))$ registers, the surplus registers can be turned into identity operators. We call these \textit{redundant} registers.

While the above provides the number of qubits that are needed to represent the commutation relations represented by $\mathcal{P}$, while still retaining Clifford equivalence, it does not immediately provide Clifford equivalent Pauli operators to those in $\mathcal{P}$ itself. For that, select a generating set for $\mathcal{P}$, $\mathcal{G}(\mathcal{P})$, then $M(\mathcal{G}(\mathcal{P}))$ will be a binary symmetric matrix of commutation values of elements from $\mathcal{P}$. Relating $M(\mathcal{G}(\mathcal{P}))$ to $M(\mathcal{D})$ will then provide a way to transform the minimal registers from $M(\mathcal{D})$ into ones that are Clifford equivalent to $\mathcal{P}$. As remarked before, composition of generators is equivalent to a simultaneous row and column addition in the commutator matrix, which forms our elementary matrix operations for a symmetric Gaussian elimination. Then there is a sequence of these operations, whose product of row operations is $L$, such that:
\begin{equation}
    M(\mathcal{G}(\mathcal{P}))=LM(\mathcal{D})L^T.
\end{equation}
This then provides a minimal register representation for the commutation relations expressed in $\mathcal{P}$ itself. This technique is efficient as each step is simply Gaussian elimination. At the end of the procedure some registers are left as identity operators which may be dropped in circuits. To determine the Clifford circuit mapping from the original Pauli operators to these minimal register representations, one may apply the results of \cite{gottesman1998heisenberg,aaronson2004improved}. The prior result provides the solution to these methods, for which these methods then find a satisfying Clifford circuits.


We may further extend the observations to condition on separably measurable registers. This leads to Theorem \ref{thm:theorem1}, 
which states that for simulating a given $\mathcal{H}$ only $\frac{1}{2}rank(M)$ qubits are required, although at a classical and repetition cost. To prove this we approach the problem through a broad symplectic linear algebraic perspective. In essence, we quotient out the isotropic subspace, stratifying by possible measurement outcomes for that space, then return to a Pauli representation with updated weights. Each of the remaining subspaces may have different coefficients, and so without further assumptions will need to be simulated.

\begin{proof}
Let $\textbf{P}$ be the non-Abelian group generated by $\{P_i\}$ with $d$ generators for $\textbf{P}$. Within $\textbf{P}$ there is a set of generators such that the commutation matrix for these generators is $\tilde{D}$, as $\textbf{P}$ contains all Gaussian eliminations. Let this set of generators with commutation matrix $\tilde{D}$ be given by $\tilde{g}$. Then the isotropic generators (those which contribute $[0]$ terms to the direct sum) are Clifford equivalent to $Z_i$ single register Paulis in the minimal register form. Upon reconstructing minimal register versions for $\{P_i\}$ from the generators $\{Z_i\}$ and $\{X_i,Z_i\}$, these $Z_i$ will still commute with these Pauli operators, and moreover qubit-wise commute. This then means that we may measure these Pauli operators and update the $P_i$ constructed by quotienting out by $Z_i$ and updating the coefficient based on the measurement value. From the direct sum decomposition, we will have $dim(M)-rank(M)$ $Z_i$ we may measure, and $rank(M)$ generators which cannot be without loss of information. This will require $\frac{1}{2}rank(M)$ qubits to represent these Paulis, and one may select not to measure all these $Z_i$, so our result is shown.
\end{proof}

It's worth noting that inherently the matrix $M(\mathcal{G} (\mathcal{P}))$, following the procedure, automatically separates out the $Z$ only registers as these registers only consist of $Z$ operators from the bases we selected for the matrix $M(\mathcal{D})$.
In the body of the work, we also describe this process from an algorithmic Clifford group perspective. This reduction is represented by the following symplectic matrix representation:
{\footnotesize{\[
\begin{tikzpicture}[]
\matrix(M)[matrix of math nodes, nodes in empty cells, nodes={anchor=center}, left delimiter=(, right delimiter=)]{
1 & 0 &   & 0 & 0 &   &   & 0 & 0 &   & 0 &&& 0 & 0 &   & 0 & 0 &   &   & 0 & 0 &   & 0 \\
* &   &   &   &   &   &   &   &   &   &   &&& * &   &   &   &   &   &   &   &   &   &   \\
  &   &   & 0 &   &   &   &   &   &   &   &&&   &   &   & 0 &   &   &   &   &   &   &   \\
* &   & * & 1 & 0 &   &   & 0 & 0 &   & 0 &&& * &   & * & 0 & 0 &   &   & 0 & 0 &   & 0 \\
0 & 0 &   & 0 & 0 &   &   & 0 & 0 &   & 0 &&& 1 & 0 &   & 0 & 0 &   &   & 0 & 0 &   & 0 \\
* &   &   &   &   &   &   &   &   &   &   &&& * &   &   &   &   &   &   &   &   &   &   \\
  &   &   & 0 &   &   &   &   &   &   &   &&&   &   &   & 0 &   &   &   &   &   &   &   \\
* &   & * & 0 & 0 &   &   & 0 & 0 &   & 0 &&& * &   & * & 1 & 0 &   &   & 0 & 0 &   & 0 \\
* &   &   & * & 0 &   &   & 0 & 0 &   & 0 &&& * &   &   & * & 1 & 0 &   & 0 & 0 &   & 0 \\
  &   &   &   &   &   &   &   &   &   &   &&&   &   &   &   & * &   &   &   &   &   &   \\
  &   &   &   &   &   &   &   &   &   &   &&&   &   &   &   &   &   &   & 0 &   &   &   \\
* &   &   & * & 0 &   &   & 0 & 0 &   & 0 &&& * &   &   & * & * &   & * & 1 & 0 &   & 0 \\
* &   &   & * & 0 &   &   & 0 & 0 &   & 0 &&& * &   &   & * & * &   &   & * & 0 &   & 0 \\
  &   &   &   &   &   &   &   &   &   &   &&&   &   &   &   &   &   &   &   &   &   &   \\
* &   &   & * & 0 &   &   & 0 & 0 &   & 0 &&& * &   &   & * & * &   &   & * & 0 &   & 0 \\
};
\draw[densely dotted](M-1-1)--(M-4-4);
\draw[densely dotted](M-2-1)--(M-4-3);
\draw[densely dotted](M-1-2)--(M-3-4);
\draw[densely dotted](M-1-2)--(M-1-4);
\draw[densely dotted](M-2-1)--(M-4-1);
\draw[densely dotted](M-1-4)--(M-3-4);
\draw[densely dotted](M-4-1)--(M-4-3);
\draw[densely dotted](M-5-1)--(M-8-4);
\draw[densely dotted](M-6-1)--(M-8-3);
\draw[densely dotted](M-5-2)--(M-7-4);
\draw[densely dotted](M-5-2)--(M-5-4);
\draw[densely dotted](M-6-1)--(M-8-1);
\draw[densely dotted](M-5-4)--(M-7-4);
\draw[densely dotted](M-8-1)--(M-8-3);
\draw[densely dotted](M-1-5)--(M-1-8);
\draw[densely dotted](M-1-9)--(M-1-11);
\draw[densely dotted](M-4-5)--(M-4-8);
\draw[densely dotted](M-4-9)--(M-4-11);
\draw[densely dotted](M-5-5)--(M-5-8);
\draw[densely dotted](M-5-9)--(M-5-11);
\draw[densely dotted](M-8-5)--(M-8-8);
\draw[densely dotted](M-8-9)--(M-8-11);
\draw[densely dotted](M-9-1)--(M-9-4);
\draw[densely dotted](M-9-5)--(M-9-8);
\draw[densely dotted](M-9-9)--(M-9-11);
\draw[densely dotted](M-12-1)--(M-12-4);
\draw[densely dotted](M-12-5)--(M-12-8);
\draw[densely dotted](M-12-9)--(M-12-11);
\draw[densely dotted](M-13-1)--(M-13-4);
\draw[densely dotted](M-13-5)--(M-13-8);
\draw[densely dotted](M-13-9)--(M-13-11);
\draw[densely dotted](M-15-1)--(M-15-4);
\draw[densely dotted](M-15-5)--(M-15-8);
\draw[densely dotted](M-15-9)--(M-15-11);
\draw[densely dotted](M-9-1)--(M-12-1);
\draw[densely dotted](M-13-1)--(M-15-1);
\draw[densely dotted](M-9-4)--(M-12-4);
\draw[densely dotted](M-13-4)--(M-15-4);
\draw[densely dotted](M-1-5)--(M-4-5);
\draw[densely dotted](M-5-5)--(M-8-5);
\draw[densely dotted](M-9-5)--(M-12-5);
\draw[densely dotted](M-13-5)--(M-15-5);
\draw[densely dotted](M-1-8)--(M-4-8);
\draw[densely dotted](M-5-8)--(M-8-8);
\draw[densely dotted](M-9-8)--(M-12-8);
\draw[densely dotted](M-13-8)--(M-15-8);
\draw[densely dotted](M-1-9)--(M-4-9);
\draw[densely dotted](M-5-9)--(M-8-9);
\draw[densely dotted](M-9-9)--(M-12-9);
\draw[densely dotted](M-13-9)--(M-15-9);
\draw[densely dotted](M-1-11)--(M-4-11);
\draw[densely dotted](M-5-11)--(M-8-11);
\draw[densely dotted](M-9-11)--(M-12-11);
\draw[densely dotted](M-13-11)--(M-15-11);
\draw[densely dotted](M-1-14)--(M-4-17);
\draw[densely dotted](M-2-14)--(M-4-16);
\draw[densely dotted](M-1-15)--(M-3-17);
\draw[densely dotted](M-1-15)--(M-1-17);
\draw[densely dotted](M-2-14)--(M-4-14);
\draw[densely dotted](M-1-17)--(M-3-17);
\draw[densely dotted](M-4-14)--(M-4-16);
\draw[densely dotted](M-5-14)--(M-8-17);
\draw[densely dotted](M-6-14)--(M-8-16);
\draw[densely dotted](M-5-15)--(M-7-17);
\draw[densely dotted](M-5-15)--(M-5-17);
\draw[densely dotted](M-6-14)--(M-8-14);
\draw[densely dotted](M-5-17)--(M-7-17);
\draw[densely dotted](M-8-14)--(M-8-16);
\draw[densely dotted](M-9-18)--(M-12-21);
\draw[densely dotted](M-10-18)--(M-12-20);
\draw[densely dotted](M-9-19)--(M-11-21);
\draw[densely dotted](M-9-19)--(M-9-21);
\draw[densely dotted](M-10-18)--(M-12-18);
\draw[densely dotted](M-9-21)--(M-11-21);
\draw[densely dotted](M-12-18)--(M-12-20);
\draw[densely dotted](M-1-18)--(M-1-21);
\draw[densely dotted](M-1-22)--(M-1-24);
\draw[densely dotted](M-4-18)--(M-4-21);
\draw[densely dotted](M-4-22)--(M-4-24);
\draw[densely dotted](M-5-18)--(M-5-21);
\draw[densely dotted](M-5-22)--(M-5-24);
\draw[densely dotted](M-8-18)--(M-8-21);
\draw[densely dotted](M-8-22)--(M-8-24);
\draw[densely dotted](M-9-14)--(M-9-17);
\draw[densely dotted](M-9-22)--(M-9-24);
\draw[densely dotted](M-12-14)--(M-12-17);
\draw[densely dotted](M-12-22)--(M-12-24);
\draw[densely dotted](M-13-14)--(M-13-17);
\draw[densely dotted](M-13-18)--(M-13-21);
\draw[densely dotted](M-13-22)--(M-13-24);
\draw[densely dotted](M-15-14)--(M-15-17);
\draw[densely dotted](M-15-18)--(M-15-21);
\draw[densely dotted](M-15-22)--(M-15-24);
\draw[densely dotted](M-9-14)--(M-12-14);
\draw[densely dotted](M-13-14)--(M-15-14);
\draw[densely dotted](M-9-17)--(M-12-17);
\draw[densely dotted](M-13-17)--(M-15-17);
\draw[densely dotted](M-1-18)--(M-4-18);
\draw[densely dotted](M-5-18)--(M-8-18);
\draw[densely dotted](M-13-18)--(M-15-18);
\draw[densely dotted](M-1-21)--(M-4-21);
\draw[densely dotted](M-5-21)--(M-8-21);
\draw[densely dotted](M-13-21)--(M-15-21);
\draw[densely dotted](M-1-22)--(M-4-22);
\draw[densely dotted](M-5-22)--(M-8-22);
\draw[densely dotted](M-9-22)--(M-12-22);
\draw[densely dotted](M-13-22)--(M-15-22);
\draw[densely dotted](M-1-24)--(M-4-24);
\draw[densely dotted](M-5-24)--(M-8-24);
\draw[densely dotted](M-9-24)--(M-12-24);
\draw[densely dotted](M-13-24)--(M-15-24);
\draw[thick]({$(M-1-12)!.5!(M-1-13)$} |- M.north)--({$(M-14-12)!.5!(M-14-13)$} |- M.south);
\draw[]({$(M.west)!.5!(M-4-1)$} |- {$(M-4-1)!.5!(M-5-1)$})--({$(M-4-11)!.5!(M-4-12)$} |- {$(M-4-1)!.5!(M-5-1)$});
\draw[]({$(M.west)!.5!(M-8-1)$} |- {$(M-8-1)!.5!(M-9-1)$})--({$(M-8-11)!.5!(M-8-12)$} |- {$(M-8-1)!.5!(M-9-1)$});
\draw[]({$(M.west)!.5!(M-12-1)$} |- {$(M-12-1)!.5!(M-13-1)$})--({$(M-12-11)!.5!(M-12-12)$} |- {$(M-12-1)!.5!(M-13-1)$});
\draw[]({$(M-1-4)!.5!(M-1-5)$} |- {$(M.north)!.5!(M-1-4)$})--({$(M-14-4)!.5!(M-14-5)$} |- {$(M.south)!.5!(M-14-4)$});
\draw[]({$(M-1-8)!.5!(M-1-9)$} |- {$(M.north)!.5!(M-1-8)$})--({$(M-14-8)!.5!(M-14-9)$} |- {$(M.south)!.5!(M-14-8)$});
\draw[]({$(M.east)!.5!(M-4-24)$} |- {$(M-4-24)!.5!(M-5-24)$})--({$(M-4-13)!.5!(M-4-14)$} |- {$(M-4-24)!.5!(M-5-24)$});
\draw[]({$(M.east)!.5!(M-8-24)$} |- {$(M-8-24)!.5!(M-9-24)$})--({$(M-8-13)!.5!(M-8-14)$} |- {$(M-8-24)!.5!(M-9-24)$});
\draw[]({$(M.east)!.5!(M-12-24)$} |- {$(M-12-24)!.5!(M-13-24)$})--({$(M-12-13)!.5!(M-12-14)$} |- {$(M-12-24)!.5!(M-13-24)$});
\draw[]({$(M-1-17)!.5!(M-1-18)$} |- {$(M.north)!.5!(M-1-17)$})--({$(M-14-17)!.5!(M-14-18)$} |- {$(M.south)!.5!(M-14-17)$});
\draw[]({$(M-1-21)!.5!(M-1-22)$} |- {$(M.north)!.5!(M-1-21)$})--({$(M-14-21)!.5!(M-14-22)$} |- {$(M.south)!.5!(M-14-21)$});
\end{tikzpicture}
\]}}
Without further assumptions on the structure of the elements of $\mathcal{P}$, no further reductions in the number of qubits needed may be made without losing some of the non-commuting properties.


Before closing out this appendix, we will briefly discuss other local-dimension cases. The term \textit{local-dimension} specifies the number of orthogonal states in each register. We will briefly discuss the case of prime power valued ($q^b$) local-dimensions.
For these cases the same mapping from Pauli operators to symplectic vectors exists \cite{gunderman2020local,ketkar2006nonbinary}. The symplectic product in this case, for vectors $s_1=(\vec{a}_1\ |\ \vec{b}_1)$ and $s_2=(\vec{a}_2\ |\ \vec{b}_2)$, is given by:
\begin{equation}
    s_1\odot s_2=\vec{a}_1\cdot \vec{b}_2-\vec{b}_1\cdot \vec{a}_2 \mod q^b.
\end{equation}
The matrix $M$ is then filled by the symplectic product of the pairs of the Pauli operators.

The dimension of $M$ is independent of the local-dimension for these cases, however, if the local-dimension is composite valued the dimension may change. The $rank(M)$ may change, however, based on the local-dimension as it is computed over the field $q^b$. Regardless the same expression holds as before since the same symmetric Gaussian elimination technique is used -- with the symplectic basis matrices in the $M(\mathcal{D})$ form replaced by $\begin{bmatrix} 0 & -1\\ 1 & 0\end{bmatrix}$. This then provides:

\begin{corollary}
Let $\mathcal{H}$ be as in Eq.~\eqref{eq:Hamiltonian_input} and $M$ the matrix of the commutation coefficients for any generating subset of the $\{ P_i \}_{i=1}^{N}$, where the Pauli operators are qudit operators with local-dimension a finite Field $\mathcal{F}$ with characteristic $q$. Then, for any choice $0\leq c \leq \dim(M)-\rank(M)$, $\mathcal{H}$ can be simulated using $\frac{1}{2}\rank(M)+[\dim(M)-\rank(M)-c]$ registers and a maximum of $q^c$ subproblems.

\end{corollary}

An important note about this result is that the rank must be computed over the field $\mathcal{F}$. Whilst qubit transformations are the primary ones that exist and are used at this time, this provides a significant generality. Additionally, if there is some method which produces collections of Pauli operators, and only depends on the commutation relations, outside of Hamiltonian simulation this would cover these cases as well.


%
\section{Physical models}
\label{app:phys_mod}
In this section, we provide additional details about the physical models employed in the plot of Fig.~\ref{fig:res}. We remark that the symmetries that our method finds is entirely independent on the values of the coefficients within the following Hamiltonians.
\begin{figure}
\centering
\includegraphics[width=\columnwidth]{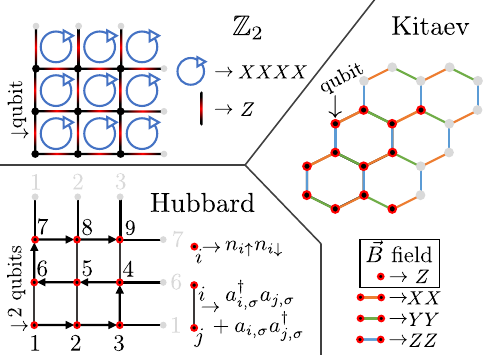}
\caption{
Physical models considered for the plot in Fig.~\ref{fig:res}. \textit{(top left)}: For the $\mathbb{Z}_2$ model \cite{Ferguson2021,chan2023hybrid,Haase2021resourceefficient,Horn1979} qubits lie on the edges of a square lattice. Each qubit individually is acted upon a $Z$ operator, and $XXXX$ operators act on plaquettes, as indicated by the blue circular arrows.
\textit{(right)}: In the Kitaev hexagonal lattice \cite{bespalova2021quantum,KITAEV20062}, qubits lie on vertices. Depending on the direction of the connection with the neighbouring vertex, the interaction is $XX$ (orange), $YY$ (green) or $ZZ$ (blue). When the magnetic field $\vec{B}$ is present, additional $Z$ operators act on each individual qubit.
\textit{(bottom left)}: In the Hubbard model \cite{stanisic2022observing}, each vertex $i$ can host two fermions with opposite spin $\sigma \in \{ \uparrow,\downarrow \}$, that interact together via $n_{i,\uparrow} n_{i,\downarrow}$. Furthermore, each fermionic species can jump between connected vertices via the flip-flop interactions. As explained in Ref.~\cite{stanisic2022observing} and in App.~\ref{app:phys_mod}, to simulate the Hubbard model with a quantum computer, fermions are encoded into qubits via the Jordan-Wigner \cite{jordan1993paulische,paulson2021} transformation, resulting in the Hamiltonian obtained from Eqs.~\eqref{eq:hubbard_ham} and \eqref{eq:hubbard_JW}
All lattices presented here are the minimal instances considered in Sec.~\ref{sec:main_res} Larger ones are obtained by increasing both dimensions equally. Furthermore, grey dots indicated periodic boundary condtions (as explicitly indicated in the Hubbard model).
}
\label{fig:phys_mod}
\end{figure}
\subsection{$\mathbb{Z}_2$ model}
The $\mathbb{Z}_2$ model \cite{Ferguson2021,chan2023hybrid,Haase2021resourceefficient,Horn1979} is a pure (without matter) lattice gauge theory (LGT) possessing a $\mathbb{Z}_2$ symmetry. As shown in Fig.~\ref{fig:phys_mod}, we consider an $N\times N$ lattice ($2\times 2$ in the figure) with qubits lying on edges. The electric field of the LGT is described by individual Pauli operators $Z$ acting over each edge. The magnetic field, on the other side, is characterized by the electric field circulating into each plaquette (i.e., the smallest cell of the lattice). The Hamiltonian is therefore
\begin{equation}
\hat{\mathcal{H}} = - \sum_{p=1}^{n_{p}}\hat{P}_{p} + \xi\sum_{q=1}^{n}\hat{Z}_{q};
\quad
\hat{P}_{p} = \prod_{i \in \square_{p}} \hat{X}_{i},
\label{eq: pcham}
\end{equation}
where $n_p$ is the number of plaquettes $\square_{p}$ ($n_p=9$ in Fig.~\ref{fig:phys_mod}) and $\xi$ is the coupling constant of the theory \cite{paulson2021}. For clarity, we included suffices to the Pauli operators to indicate to which qubit they act (e.g., $X_3$ indicates $X$ acts on the third qubit).

\subsection{Kitaev honeycomb model}
As shown in Fig.~\ref{fig:phys_mod}, the Kitaev honeycomb model employs a hexagonal lattice on a torus. Following Ref.~\cite{bespalova2021quantum}, the Hamiltonian of the system is
\begin{equation}\label{eq:kitaev}
    \begin{split}
        \mathcal{H} = & J_{\rm x}\sum_{\langle i,j \rangle \in \mathcal{X}} X_i X_j
        +
        J_{\rm y}\sum_{\langle i,j \rangle \in \mathcal{Y}} Y_i Y_j
        \\ 
        & +
        J_{\rm z}\sum_{\langle i,j \rangle \in \mathcal{Z}} Z_i Z_j
        +
        \sum_{i=1}^{n} h_{\text{z},i} Z_i,
    \end{split}
\end{equation}
where $\mathcal{X}$, $\mathcal{Y}$, $\mathcal{Z}$ refer to edges aligned to the three defining directions of an hexagon: in the figure, the orange, green and blue, respectively. The last term in Eq.~\eqref{eq:kitaev} describes $\vec{B}$, such that when all coefficients $h_{\text{z},i}$ are equal to (different from) zero, the model is with (without) magnetic field. The remaining coefficients $J_{\rm x}$, $J_{\rm y}$ and $J_{\rm z}$ characterize the neighbour-neighbour interactions of the model.

\subsection{Hubbard model}
As another concrete example we consider the Hubbard model on $n/2$ sites (such that the number of qubits is $n$). In this case there is an analytic solution for the dimension of the isotropic space, and hence the parallelization size. The Hamiltonian is given by
\begin{equation}\label{eq:hubbard_ham}
    H=-\sum_{\substack{\langle i,j\rangle \\ \sigma\in\{\uparrow,\downarrow\}}}\left( a_{i\sigma}^\dag a_{j\sigma}+a_{j\sigma}^\dag a_{i\sigma}\right)+U\sum_{i=1}^{n/2} n_{i\uparrow}n_{i\downarrow},
\end{equation}
where these operators describe two species ($\sigma \in \{ \uparrow,\downarrow \}$) of fermions residing at each lattice site (see Fig.~\ref{fig:phys_mod} and Ref.~\cite{stanisic2022observing}). The first (second) summation describes interactions between fermions of the same species residing on neighbouring vertices $i,j$ (fermions of different species residing on the same vertex $i$). $U$ is a parameter describing the theory.

We will assume that there is a single connected lattice. If there is more than one lattice then the savings will be the resultant amount for each lattice. Beyond this, there are no restrictions placed on the connectivity of the spins. This Hamiltonian is translated into qubit operators via the Jordan-Wigner transformation \cite{jordan1993paulische}
\begin{subequations}\label{eq:hubbard_JW}
\begin{align}
    \left( a_{i\sigma}^\dag a_{j\sigma}+a_{j\sigma}^\dag a_{i\sigma}\right)
    & \mapsto &
    (X_iX_j+Y_iY_j)\bigotimes_{k=i+1}^{j-1} Z_k, 
    \\
    n_{i\uparrow}n_{i\downarrow}
    & \mapsto &
    \frac{(I-Z_{i\uparrow})(I-Z_{i\downarrow})}{4},
    \end{align}
\end{subequations}
where the numbering follows the ``snake pattern'' shown in Fig.~\ref{fig:phys_mod} and described in Ref.~\cite{stanisic2022observing}.
Since we only need a compositionally independent set the latter equation generates $Z$ operators on each register in the $\uparrow$ and $\downarrow$ lattices. Then the excitation preserving operators may have their $Z$ operator terms removed without changing our final result. Following this, for each $X_iX_j$ we may remove the corresponding $Y_iY_j$ operator since these only differ by $Z$ operators.

Next, we may cancel out some of the $X_iX_j$ terms since they are not all independent. For any loop the $X_iX_j$ terms cancel out, and so we may remove a generator to break the loop and all terms will then be independent. Since this holds for all loops, we may repeat this process until the connectivity is a tree of $n$ vertices, and so $n-1$ independent $X_iX_j$ Pauli operators will exist for each of the spin lattices. The total number of independent Pauli operators is then $n-2+n$.

To find the rank of the commutation matrix, we first note that the spin direction lattices are uncoupled for this computation. The simplest way to find the rank in this case is then to note that for each $Z$ Pauli operator there are a unique pair of $X_iX_j$ operators which do not commute with it, and the converse is also true. Then the commutation matrix will be as full rank as possible, which given that the dimension is odd means that the rank will be $dim(M)-1$, and so the isotropic space is of size $1$ for each of the two spin direction lattices. Therefore the total isotropic subspace size is $2$, meaning that $2$ qubits may be saved through this technique.

\section{Interpretation within Quantum Error-Correcting Codes}
\label{app:QECC}

While these observations and statements do not impact the results of this work, they do help place it within the context of quantum error-correcting codes and may spur further advances in these procedures. The routine referred to as $\mathcal{RR}$ in Sec.~\ref{sec:main_res}, as noted in \cite{gunderman2023transforming}, has a clear connection with Entanglement-Assisted Quantum Error-Correcting Codes (EAQECC) and Quantum Convolutional Codes. In EAQECC the $\mathcal{RR}$ routine corresponds to finding the minimal number of priorly shared entangled bits that must be shared, albeit in EAQECC generators can be freely selected whereas here that is not permitted \cite{brun2006correcting,hsieh2008entanglement,wilde2008optimal}. In quantum convolutional codes the same problem occurs in finding the minimal number of qubits that must be carried between frames in the code \cite{wilde2011examples,houshmand2012minimal}. These are both primarily communication settings, whereas the $\mathcal{RR}$ method is set within Pauli reductions and only permits Clifford conjugations.

While those prior connections somewhat naturally occur, the routine $\mathcal{CR}$ also can be cast within a quantum error-correcting code setting. In this case, the removal of the isotropic subspace corresponds to removing the stabilizer generators from a subsystem quantum error-correcting code and only working with the Gauge operators \cite{kribs2005unified,poulin2005stabilizer}. While this connection is not utilized in this work the efforts within subsystem quantum error-correcting codes works to reduce the stabilizer weights could prove useful in pursuing a different avenue for this line of work \cite{bravyi2012subsystem,higgott2021subsystem}. It is worth being clear, however, that while this connection exists it is not a two-way path: putting a subsystem code through this procedure will typically reduce the distance of that code due to the entangling gates in the Clifford circuit.

\section{Pseudocode}
\label{app:pseudo}
The pseudocode employed by our algorithm is the following:
\begin{lstlisting}
q_I,q_Z,q_X = [],[],[]
for i in [n,n-1,...,1]:
  if all Paulis have I on qubit i:
    add i to q_I
  else:
    pick one Pauli, P, with P[i]=X/Y/Z
    map P[1,...,i] to I...IZ
    if all Paulis have I/Z on qubit i:
      add i to q_Z
    else:
      pick one Pauli, Q, with Q[i]=X/Y
      map P[1,...,i] to I...IX
      add i to q_X
\end{lstlisting}
Using SWAP gates, we can permute the qubits such that $q_I = [1,\dotsc,r]$, $q_Z = [r+1,\dotsc,r+c]$, and $q_X = [r+c+1,\dotsc,n]$. By re-ordering the Paulis, we can make sure the first $n-r-c$ Paulis are those called $Q$ in the algorithm, and the next $n-r$ Paulis are those called $P$ in the algorithm. After these permutations, the resulting string representation is:
\[
\begin{tikzpicture}[]
\matrix(M)[matrix of nodes, font=\ttfamily, nodes in empty cells, nodes={anchor=center}]{
I &   & I & I &   &   & I & I &   & I & X \\
  &   &   &   &   &   &   &   &   &   & * \\
  &   &   &   &   &   &   & I &   &   &   \\
I &   & I & I &   &   & I & X & * &   & * \\
I &   & I & I &   &   & I & I &   & I & Z \\
  &   &   &   &   &   &   &   &   &   & * \\
  &   &   &   &   &   &   & I &   &   &   \\
I &   & I & I &   &   & I & Z & * &   & * \\
I &   & I & I &   & I & Z & * &   &   & * \\
  &   &   &   &   &   & + &   &   &   &   \\
  &   &   & I &   &   &   &   &   &   &   \\
I &   & I & Z & + &   & + & * &   &   & * \\
I &   & I & + &   &   & + & * &   &   & * \\
  &   &   &   &   &   &   &   &   &   &   \\
I &   & I & + &   &   & + & * &   &   & * \\
};
\draw[densely dotted](M-3-8)--(M-1-10);
\draw[densely dotted](M-4-8)--(M-1-11);
\draw[densely dotted](M-4-9)--(M-2-11);
\draw[densely dotted](M-7-8)--(M-5-10);
\draw[densely dotted](M-8-8)--(M-5-11);
\draw[densely dotted](M-8-9)--(M-6-11);
\draw[densely dotted](M-11-4)--(M-9-6);
\draw[densely dotted](M-12-4)--(M-9-7);
\draw[densely dotted](M-12-5)--(M-10-7);

\draw[densely dotted](M-1-1)--(M-4-1);
\draw[densely dotted](M-1-3)--(M-4-3);
\draw[densely dotted](M-1-4)--(M-4-4);
\draw[densely dotted](M-1-7)--(M-4-7);
\draw[densely dotted](M-1-8)--(M-3-8);
\draw[densely dotted](M-2-11)--(M-4-11);

\draw[densely dotted](M-5-1)--(M-8-1);
\draw[densely dotted](M-5-3)--(M-8-3);
\draw[densely dotted](M-5-4)--(M-8-4);
\draw[densely dotted](M-5-7)--(M-8-7);
\draw[densely dotted](M-5-8)--(M-7-8);
\draw[densely dotted](M-6-11)--(M-8-11);

\draw[densely dotted](M-9-1)--(M-12-1);
\draw[densely dotted](M-9-3)--(M-12-3);
\draw[densely dotted](M-9-4)--(M-11-4);
\draw[densely dotted](M-10-7)--(M-12-7);
\draw[densely dotted](M-9-8)--(M-12-8);
\draw[densely dotted](M-9-11)--(M-12-11);

\draw[densely dotted](M-13-1)--(M-15-1);
\draw[densely dotted](M-13-3)--(M-15-3);
\draw[densely dotted](M-13-4)--(M-15-4);
\draw[densely dotted](M-13-7)--(M-15-7);
\draw[densely dotted](M-13-8)--(M-15-8);
\draw[densely dotted](M-13-11)--(M-15-11);

\draw[densely dotted](M-1-1)--(M-1-3);
\draw[densely dotted](M-4-1)--(M-4-3);
\draw[densely dotted](M-5-1)--(M-5-3);
\draw[densely dotted](M-8-1)--(M-8-3);
\draw[densely dotted](M-9-1)--(M-9-3);
\draw[densely dotted](M-12-1)--(M-12-3);
\draw[densely dotted](M-13-1)--(M-13-3);
\draw[densely dotted](M-15-1)--(M-15-3);

\draw[densely dotted](M-1-4)--(M-1-7);
\draw[densely dotted](M-4-4)--(M-4-7);
\draw[densely dotted](M-5-4)--(M-5-7);
\draw[densely dotted](M-8-4)--(M-8-7);
\draw[densely dotted](M-9-4)--(M-9-6);
\draw[densely dotted](M-12-5)--(M-12-7);
\draw[densely dotted](M-13-4)--(M-13-7);
\draw[densely dotted](M-15-4)--(M-15-7);

\draw[densely dotted](M-1-8)--(M-1-10);
\draw[densely dotted](M-4-9)--(M-4-11);
\draw[densely dotted](M-5-8)--(M-5-10);
\draw[densely dotted](M-8-9)--(M-8-11);
\draw[densely dotted](M-9-8)--(M-9-11);
\draw[densely dotted](M-12-8)--(M-12-11);
\draw[densely dotted](M-13-8)--(M-13-11);
\draw[densely dotted](M-15-8)--(M-15-11);

\draw[]({$(M.west)!.5!(M-4-1)$} |- {$(M-4-1)!.5!(M-5-1)$})--({$(M-4-11)!.5!(M.east)$} |- {$(M-4-1)!.5!(M-5-1)$});
\draw[]({$(M.west)!.5!(M-8-1)$} |- {$(M-8-1)!.5!(M-9-1)$})--({$(M-8-11)!.5!(M.east)$} |- {$(M-8-1)!.5!(M-9-1)$});
\draw[]({$(M.west)!.5!(M-12-1)$} |- {$(M-12-1)!.5!(M-13-1)$})--({$(M-12-11)!.5!(M.east)$} |- {$(M-12-1)!.5!(M-13-1)$});

\end{tikzpicture}
\]
where the asterisks can be any Pauli, $I$, $X$, $Y$, or $Z$, and the pluses can be either $I$ or $Z$.

\begin{widetext}
    \subsection{Table of Results}

\begin{table}[!h]
\begin{tabular}{ |p{8.2cm}||p{2.5cm}|p{3.6cm}|p{2.6cm}|  }
 \hline
 \multicolumn{4}{|c|}{Maximal Reductions From Algorithm} \\
 \hline
 Hamiltonian & Qubits a Priori & Separable Measurements & Qubits Required\\
 \hline
 $H_2$   & $4$    & $3$ &   $1$\\
 $LiH$ & $12$ & $4$ & $8$\\
 $BeH_2$ & $14$ & $5$ & $9$\\ 
  $\mathbb{Z}_2$ LGT with Gaussian Perturbation: $2\times 2$ &   $8$  & $5$   & $3$\\
 $\mathbb{Z}_2$ LGT with Gaussian Perturbation: $5\times 5$ &   $50$  & $26$   & $24$\\
 $\mathbb{Z}_2$ LGT with Gaussian Perturbation: $6\times 6$ & $72$ & $37$ & $35$\\
 $\mathbb{Z}_2$ LGT with Gaussian Perturbation: $7\times 7$ & $98$ & $50$ & $48$ \\
 $\mathbb{Z}_2$ LGT with Gaussian Perturbation: $10\times 10$ & $200$ & $101$ & $99$ \\
 $\mathbb{Z}_2$ LGT with Gaussian Perturbation: $15\times 15$ & $450$ & $226$ & $224$ \\
 Hubbard & $2n$ & $2$ & $2(n-1)$\\
 Kitaev Honeycomb: $1\times 1$ & $4$ & $3$ & $1$\\
 Kitaev Honeycomb: $2\times 2$ & $12$ & $6$ & $6$\\
 Kitaev Honeycomb: $5\times 5$ & $60$ & $25$ & $35$\\
 Kitaev Honeycomb: $10\times 10$ & $220$ & $98$ & $122$\\
 Kitaev Honeycomb: $15\times 15$ & $480$ & $220$ & $260$\\
 Kitaev Honeycomb with $J_z$ and $J_{zz}$ Perturbation: $1\times 1$ & $4$ & $2$ & $2$\\
Kitaev Honeycomb with $J_z$ and $J_{zz}$ Perturbation: $2\times 2$ & $12$ & 3 & $9$\\
Kitaev Honeycomb with $J_z$ and $J_{zz}$ Perturbation: $5\times 5$ & $60$ & $6$ & $54$\\
Kitaev Honeycomb with $J_z$ and $J_{zz}$ Perturbation: $10\times 10$ & $220$ & $11$ & $209$\\
Kitaev Honeycomb with $J_z$ and $J_{zz}$ Perturbation: $15\times 15$ & $480$ & $16$ & $464$\\
 Kitaev Honeycomb with $J_{zz}$ Perturbation: $1\times 1$ & $4$ & $3$ & $1$\\
 Kitaev Honeycomb with $J_{zz}$ Perturbation: $2\times 2$ & $12$ & $6$ & $6$\\
  Kitaev Honeycomb with $J_{zz}$ Perturbation: $5\times 5$ & $60$ & $25$ & $35$\\
   Kitaev Honeycomb with $J_{zz}$ Perturbation: $10\times 10$ & $220$ & $98$ & $122$\\
    Kitaev Honeycomb with $J_{zz}$ Perturbation: $15\times 15$ & $480$ & $220$ & $260$\\
 
 \hline
 
\end{tabular}
\caption{Qubit requirement reductions by applying our methods. LGT with or without Gaussian perturbation have the same requirements. More work is required to determine which models benefit most from the techniques shown in this work, and there is further room for improvements beyond the methods list here, however, given the tools stated, these are optimal.}
\label{summaryresults}
\end{table} 
\end{widetext}



\end{document}